\documentclass{PoS}

\usepackage{graphicx}

\title{Narrow-Line AGNs: confirming the relationship between metallicity and accretion rate}

\ShortTitle{Narrow-Line AGNs}

\author{\speaker{D. M. Neri-Larios}\\
        Departamento de Astronom\'ia, Universidad de Guanajuato, 36000, Guanajuato, Mexico\\
        E-mail: \email{daniel@astro.ugto.mx}}

\author{R. Coziol$^1$, J.P. Torres-Papaqui$^1$, H. Andernach$^1$, J.M. Islas-Islas$^1$, I. Plauchu-Frayn$^2$, and R.A. Ortega-Minakata$^1$\\
        1 Departamento de Astronom\'ia, Universidad de Guanajuato, 36000, Guanajuato, Mexico\\
        2 Instituto de Astrof\'isica de Andaluc\'ia (CSIC), E-18008, Granada, Spain\\
        E-mail: \email{rcoziol@astro.ugto.mx, papaqui@astro.ugto.mx, heinz@astro.ugto.mx, jmislas@astro.ugto.mx, ilse@iaa.es, rene@astro.ugto.mx}}

\abstract{We have selected a sample of 292 SDSS Narrow-Emission-Line galaxies (NELGs) known to have formed and evolved in relative isolation to study the nature and origin of the AGN phenomenon. The galaxies in our sample have line fluxes with S/N$> 3$ and were separated using a standard diagnostic diagram into Star Forming Galaxies (SFGs; 36.0\%), Transition type Objects (TOs; 28.4\%) and Narrow-Line AGNs (NLAGNs; 35.6\%). Having found a strong correlation between the bulge mass and the NLAGN phenomenon, we have applied the same relation as for the Broad-Line AGNs (BLAGNs) to estimate their black hole (BH) masses. The BH in the NLAGNs are 2 to 3 orders lower in mass than the BHs found in BLAGNs, but are comparable to those observed in Narrow-Line Seyfert~1 (NLS1), although none of our objects can be classified as such. To determine the metallicities, [O/H], of the NLAGNs we  calibrated the standard diagnostic diagram [OIII]/H$\beta$ vs. [NII]/H$\alpha$ using similar relation as for the SFGs, which reproduce the values obtained with CLOUDY simulations developed for Bennert et al. [3]. For some individual objects we compared our line ratios with other CLOUDY similations by different authors. This suggests we achieve a typical uncertainty of 0.2 dex on [O/H], increasing to 0.3-0.5 in the Seyfert~2 (S2). This calibration suggests the metallicities of the NLAGNs are subsolar, varying between 1 and 0.3 $Z_\odot$. We find two statistically significant positive correlations: for [O/H] with the BH mass and for [O/H] with the luminosity at 5100\AA, $\lambda L(5100$\AA$)$. No correlation is found between [O/H] and the accretion rate, $L_{bol}/L_{Edd}$. However, comparisons with the BLAGNs suggest the NLAGNs extend the metallicity-accretion rate relationship [24] to the low metallicity regime. Although the NLS1 have similar BH masses as the NLAGNs they show higher accretion rates, which is consistent with their higher metallicities.}

\FullConference{Narrow-Line Seyfert 1 Galaxies and their place in the Universe - NLS1,\\
        April 04-06, 2011\\
        Milan Italy}

\begin{document}

\section{Introduction}
Spectroscopic surveys like the Sloan Digital Sky Survey (SDSS), have
revealed that a very high number of galaxies in the nearby universe
show narrow emission lines in their nucleus. Using spectroscopic
diagnostic diagrams comparing various line ratios, different
classification schemes were proposed to distinguish between the
possible sources of excitation of the gas in these galaxies ([2];
[28]; [15]; [13]). These classification schemes suggest there are
two main sources of ionization: thermal sources, which are related
to star forming activity, and non thermal sources that are related
to the possible accretion of matter by a massive black hole, the
so-called Active Galactic Nuclei (AGNs).

Although most of the emission line galaxies in SDSS are ionized by
stars, recent studies (e.g, [20]; [27]) suggest that the actual
number of galaxies with an accreting black hole may equal that of
the star forming galaxies. However, the evidence is obscured by the
large variation of characteristics observed in AGNs. For example, in
the so-called Seyfert~1 (S1) galaxies, where we distinguish broad
emission line components akin to what is observed in quasars ([21];
[29]; [18]), the accretion of matter onto a supermassive BH seems
obvious. But the situation is much more complicated with narrow-line
AGNs (NLAGNs) where no broad emission lines are observed. We usually
distinguish two kinds of NLAGNs: the high-ionization Seyfert~2 (S2)
and the low-ionization LINER, which stands for Low Ionization
Nuclear Emission-line Region ([10]; [5]; [14]). Another case which
is particularly intriguing is the narrow-line Seyfert~1 (NLS1),
which seems to show the same high level of activity as the S1 but
with narrower emission lines ([17]; [16]).

Understanding what is the nature of the differences between the
various NLAGNs is an important matter and the subject of intense
research activity. According to the unification model for AGNs, the
difference between BLAGNs and NLAGNs is solely a question of
geometry and orientation with respect to the line of sight: all
these galaxies have a BH at their center, but in the NLAGNs the line
of sight to the region producing the broad emission lines is
obscured by a torus of matter and dust encircling the BH. However,
the prevalent model to explain the NLS1 seems to suggest they have
smaller mass BHs than the S1, and, like quasars, accrete matter near
the Eddington limit [23]. But, what happens then for the S2 and
LINER? Here we use a new sample of isolated galaxies classified as
NLAGNs to study this problem.

\section{Description of the sample and determination of activity type}

\begin{figure}[h!]
\centering
\includegraphics[width=11cm,height=10.5cm]{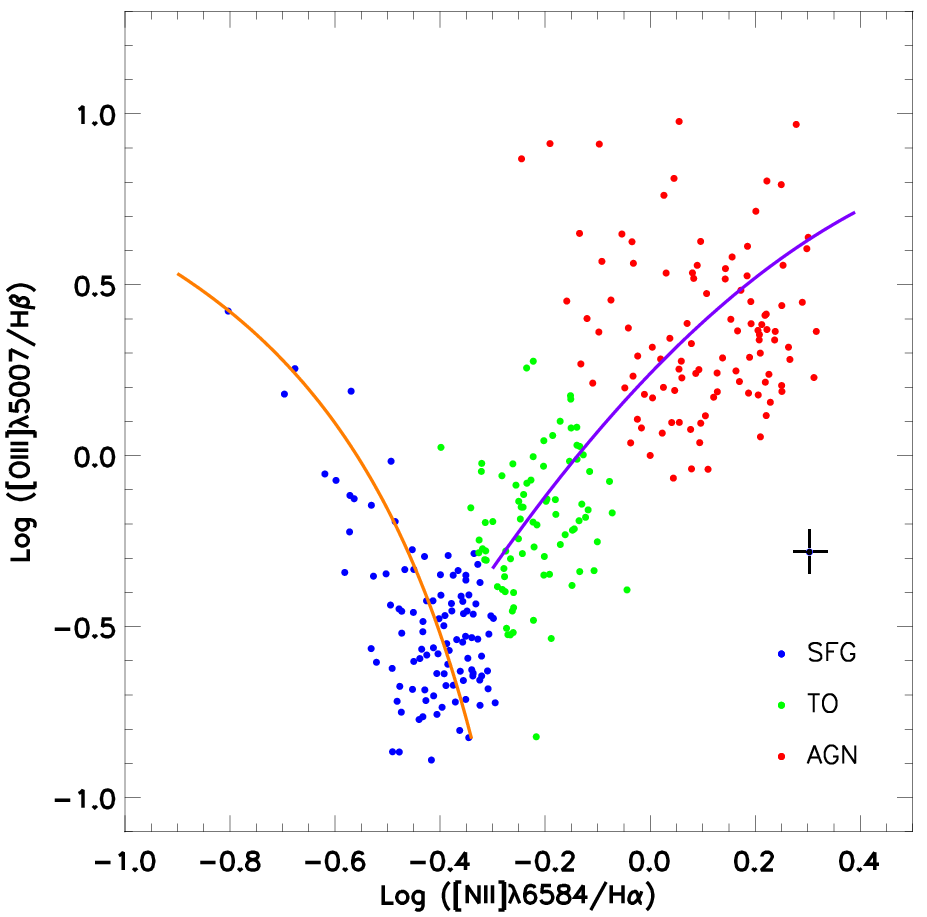}
\caption{Metallicity calibrated diagnostic diagram. The inversion of
the relation between [O/H] and [NII]/H$\alpha$ for the AGNs explains
the $\nu$ feature of the standard diagnostic diagram.} \label{fig:1}
\end{figure}

By comparing the position of galaxies in the 2MASS-selected Isolated
Galaxies Catalog (2MIG; [11]) with those in the Sloan Digital Sky
Survey Data Release 7 (SDSS DR7, [1]) we have retrieved the spectra
for 292 narrow emission line galaxies, with signal-to-noise ratios
S/N $\ge$ 3.0 for [OIII]$\lambda$5007, H$\beta$, [NII]$\lambda$6584
and H$\alpha$. The spectra were subsequently corrected for Galactic
extinction, shifted to their rest frame, re-sampled to $\Delta
\lambda$ = 1\AA\ between 3400 and 8900\AA, and processed using the
spectral synthesis code STARLIGHT [4]. Using STARLIGHT the
contribution of the underlying stellar population was effectively
eliminated, allowing a high precision measure of the fluxes of the
different emission lines observed. Through the template fitting
technique, we also obtained an estimate of the stellar velocity
dispersion in the bulge of the galaxies.

The activity types of the galaxies were determined using the
standard diagnostic diagram, Figure~\ref{fig:1}, that compares the
ratios [OIII]$\lambda$5007/ H$\beta$ and
[NII]$\lambda$6584/H$\alpha$. The typical low uncertainty levels on
these ratios (illustrated by the cross in Figure~\ref{fig:1}) have
practically no effect on our classification. We distinguish 105 SFG
galaxies (36.0\%), 83 TOs (28.4\%), and 104 AGNs (35.6\%). Note that
we do not distinguish between S2 and LINERs in our sample, simply
because of the lack of clear distinction in this diagnostic diagram
between these two types of activity.

\section{Black Hole masses, luminosities and metallicities of the NLAGNs}

\begin{figure}[h!]
\centering
\includegraphics[width=16cm,height=10cm]{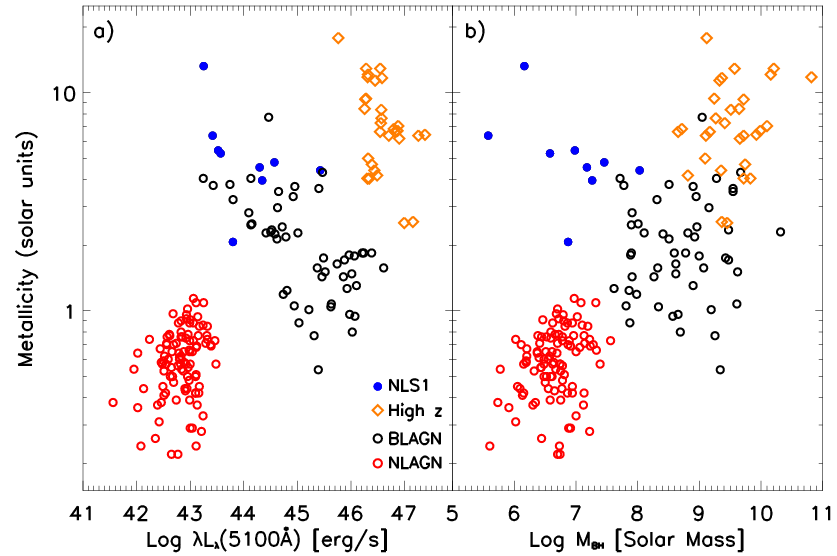}
\caption{a) Metallicities in solar units, Z/Z$_\odot$, versus
5100\AA\ luminosities, and b) versus Black hole masses.}
\label{fig:3}
\end{figure}

In two recent articles, Shemmer et al. [24]  and Matsuoka et al.
[19], it was shown that in BLAGNs the continuum luminosity at
5100\AA\ and the black hole mass are correlated with the
metallicity. The NLS1, however, seem to distinguish themselves by
showing lower black hole masses for their high metallicities (see
Figure~\ref{fig:3}b). However, if one transforms the luminosities
into accretion rates, the NLS1 seem to follow the same relation as
the BLAGNs (see Figure~\ref{fig:4}).

\begin{figure}[t!]
\centering
\includegraphics[width=10cm,height=10cm]{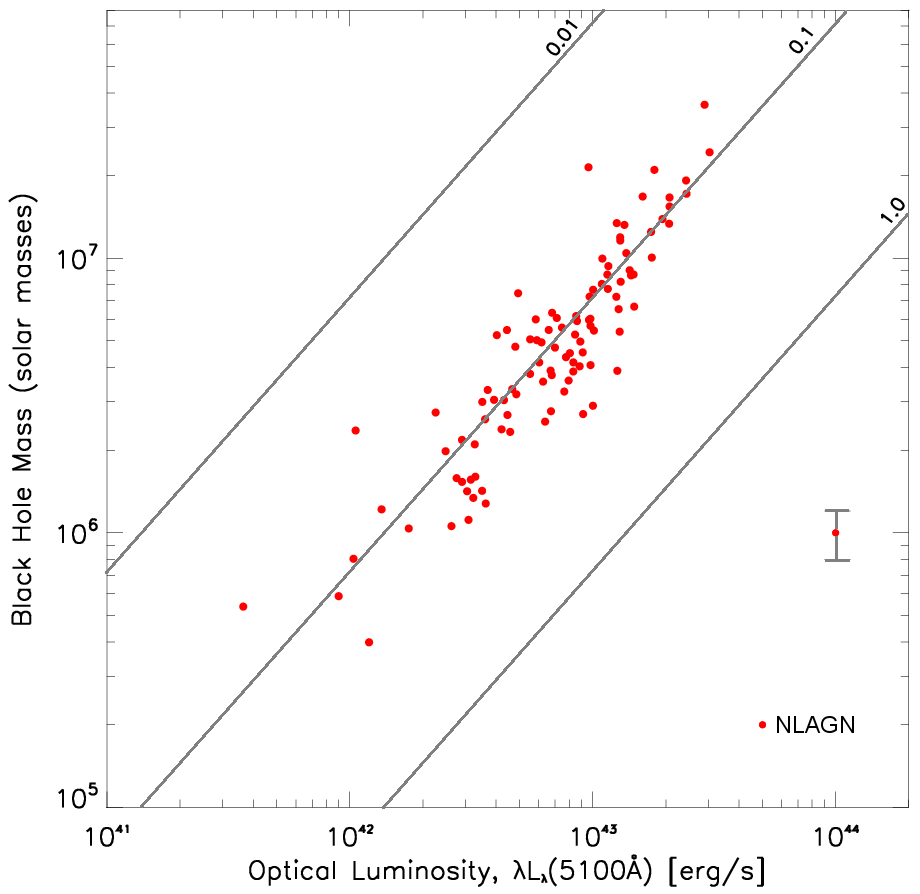}
\caption{Black hole mass as a function of 5100\AA\  luminosity for
the NLAGNs. The typical uncertainty on the BH masses is indicated by
a red dot with error bar in the lower right part of the figure.}
\label{fig:2}
\end{figure}

To verify if this behavior is also observed in the 104 NLAGNs in our
sample, we have determined their BH masses, luminosities and
estimated their metallicities [6]. The BH masses were obtained from
the bulge masses, assuming the same relation as for the BLAGNs,
proposed by H\"aring \& Rix [9]. The $\lambda L(5100$\AA$)$ was
measured directly in the SDSS spectra before subtracting the stellar
population templates. We have verified that our method yields
luminosities at 5100\AA\ similar (one-to-one relation) to those
obtained by fitting a power law directly on this region of the
spectra [7]. In Figure~\ref{fig:2} we compare the BH masses with the
luminosity $\lambda L(5100$\AA$)$. The NLAGNs have BH masses much
below 10$^8$ M$_{\odot}$, which is the lower limit in BLAGNs. These
masses are comparable to those measured in the NLS1 (Figure
\ref{fig:3}b).

\begin{table}
\begin{center}
\caption{Pearson ($r$) and Spearman ($r_s$) correlation tests and chance probabilities.}
\label{tab:1}
\begin{tabular}{lcccc}
\hline
[O/H] versus & $r$ & $P(r)$ & $r_s$ & $P(r_s)$\\
\hline
$\lambda L_\lambda(5100\AA)$ &   0.28  &   0.0037  &  0.28    & 0.0045  \\
BH mass                   &   0.28  &   0.0035  &  0.30    & 0.0028   \\
$L_{bol}/L_{Edd}$         &  -0.05  &   0.5932  & -0.09    & 0.3511   \\
\hline
\end{tabular}
\end{center}
\end{table}

The metallicity, [O/H], was  determined empirically by
fitting--purple line in Figure \ref{fig:1}--a correlation between
OIII]$\lambda$5007/H$_\beta$ and [NII]/H$\alpha$ on the TOs,
extrapolating over the AGNs, and by searching for a metallicity
calibration that reproduces the values obtained by Bennert et al.
[3] using {\scriptsize CLOUDY}. In Coziol et al. [6] we found that
the calibration based on $R_3$ =
1.35$\times$([OIII]$\lambda$5007/H$_\beta$) as determined
empirically for the SFGs yields a relatively good first
approximation, with uncertainty of $\pm 0.2$ dex, but slightly
higher for the S2 which are extreme objects compared to most of the
AGNs in our sample. In Figure~\ref{fig:1} the metallicity
calibration for the NLAGNs region implies the ratio [NII]/H$\alpha$
increases as [O/H] decreases through the relation: [O/H] = $-$0.52 +
($\log$([NII]/H$\alpha$) $-$ 0.6)$^2$. This is consistent with the
possibility of an overabundance of nitrogen in AGNs ([22]; [26];
[25]; [8]; [3]).

In Figure \ref{fig:3} we compare the metallicities of the NLAGNs
with their luminosities $\lambda L(5100$\AA$)$ and BH masses. We
find two statistically significant positive correlations, for [O/H]
with $\lambda L(5100$\AA$)$ and [O/H] with the BH mass (see Table
\ref{tab:1}).

In Figure \ref{fig:3} we also compare the NLAGNs with the BLAGNs at
low and high z and the NLS1, as studied before by Shemmer et al.
[24]. The transformation from N{\scriptsize V}/C{\scriptsize IV} to
[O/H] is based on the results of the model published by Hamann \&
Ferland [8]. We have deduced the relation $Z_{\odot}$ = 0.126 +
8.859$\cdot$(N{\scriptsize V}/C{\scriptsize IV}) -
0.949$\cdot$(N{\scriptsize V}/C{\scriptsize IV})$^2$ . Note how well
the NLAGNs in our sample seem to extend the relationships found for
the BLAGNs towards the lower regime in [O/H].

\section{The relation between metallicities and accretion rates}

\begin{figure}[h!]
\centering
\includegraphics[width=10.5cm,height=10cm]{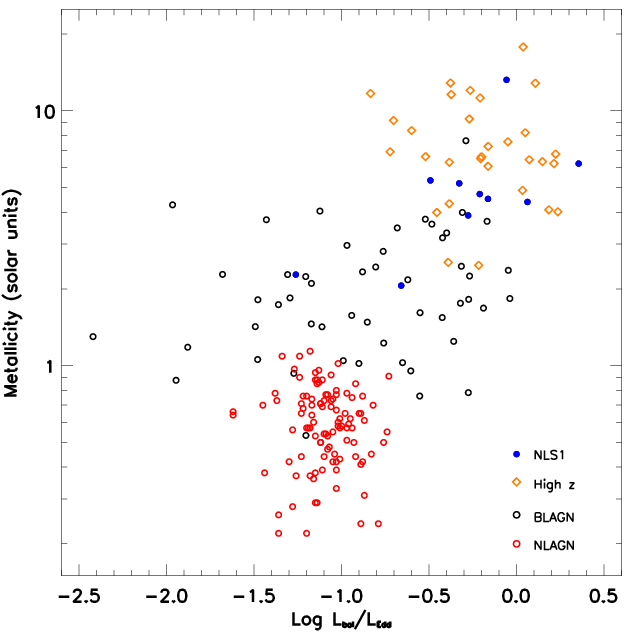}
\caption{Metallicities in solar units versus accretion rates, $L_{bol}/L_{Edd}$.}
\label{fig:4}
\end{figure}

To determine the bolometric luminosity, $L_{bol}$, of the NLAGNs, we
used the relation $L_{bol} = 9\times \lambda L(5100$\AA$)$ [11].
Although our sample of NLAGNs considered alone shows no correlation
between [O/H] and the accretion rate $L_{bol}/L_{Edd}$ (Table
\ref{tab:1}),  added to the sample of Shemmer et al. [24],
Figure~\ref{fig:4}, they seem to extend the relation found for the
BLAGNs to lower metallicities.

In their analysis, Shemmer et al. [24] also found the NLS1 to have
accretion rates comparable to BLAGNs, despite their lower BH masses,
but consistent with their high metallicities. In Figure~\ref{fig:4}
the NLAGNs which have lower metallicities than the NLS1 accrete at
lower rates, which is consistent with the metallicity-accretion rate
relationship found by Shemmer et al. [24].

\section{Conclusions}

Based on our study, we formulate four conclusions:

\begin{enumerate}
\item
The NLAGNs composed of S2 and LINERs are genuine AGNs. They simply
have smaller mass BHs than the BLAGNs. They also show lower
accretion rates, typically accreting matter at 0.1 the Eddington
limit.

\item
The metallicities of the NLAGNs are subsolar, varying between 0.3
and 1 $Z_{\odot}$. Similarly to the BLAGNs, the NLAGNs show
statistically significant positive correlations for the metallicity
with the BH mass and the luminosity at 5100\AA.

\item
The correlations observed for the NLAGNs seem to extend the
correlations of the BLAGNs to the lower metallicity regime ([24];
[19]). In particular, we find a strong correlation between the black
hole mass and metallicity, suggesting a common behavior for AGNs
([8]; [19]; [6]).

\item
The NLAGNs is consistent with the relationship between the
metallicity and accretion rate as previously discovered by Shemmer
et al. [24] for the BLAGNs. The NLAGNs have BH masses similar to
those of NLS1, but because their metallicities are lower, they
accrete at lower rates.

\end{enumerate}

\acknowledgments J.P. Torres-Papaqui acknowledges PROMEP  for
support grants 103.5-10-4684. I. Plauchu-Frayn acknowledges
postdoctoral fellowship No. 145727 from CONACyT, M\'exico. The
authors would also like to thank N. Bennert and S. Komossa for
making their results of CLOUDY models available to us, and O.
Shemmer for furnishing the data for the BLAGNs.


\begin{thebibliography}{99}
\bibitem{1} K.N. Abazajian, J.K. Adelman-McCarthy, M.A. Ag\"ueros, and 201 coauthors, 2009, \emph{The Seventh Data Release of the Sloan Digital Sky Survey}, \emph{ApJS}, {\bf 182}, 543.
\bibitem{2} J.A. Baldwin, M.M. Phillips, \& R. Terlevich, 1981, \emph{Classification parameters for the emission-line spectra of extragalactic objects}, \emph{PASP}, {\bf 93}, 5.
\bibitem{3} N. Bennert, B. Jungwiert, S. Komossa, M. Haas, \& R. Chini, 2006, \emph{Size and properties of the NLR in the Seyfert-2 galaxy NGC 1386}, \emph{A\&A}, {\bf 446}, 919.
\bibitem{4} R. Cid Fernandes, A. Mateus, L. Sodr\'e, G. Stas\'inska, \& J.M. Gomes, 2005, \emph{Semi-empirical analysis of Sloan Digital Sky Survey galaxies - I. Spectral synthesis method}, \emph{MNRAS}, {\bf 358}, 363.
\bibitem{5} R. Coziol, 1996, \emph{The history of star formation of starburst galaxies}, \emph{A\&A}, {\bf 309}, 345.
\bibitem{6} R. Coziol, J.P. Torres-Papaqui, I. Plauchu-Frayn, J.M. Islas-Islas, R.A. Ortega-Minakata, D.M. Neri-Larios, \& H. Andernach, 2011, \emph{The nature and origin of narrow-line AGN activity in galaxies}, \emph{MNRAS}, submitted.
\bibitem{ } P.J. Francis, P.C. Hewett, C.B. Foltz, F.H. Chaffee, R.J. Weymann, S.L. Morris, 1991, \emph{A high signal-to-noise ratio composite quasar spectrum}, \emph{ApJ}, {\bf 373}, 465.
\bibitem{7} F. Hamann, \& G. Ferland, 1993, \emph{The Chemical Evolution of QSOs and the Implications for Cosmology and Galaxy Formation}, \emph{ApJ}, {\bf 418}, 11.
\bibitem{8} N. H\"aring \&  H.W. Rix, 2004, \emph{On the Black Hole Mass-Bulge Mass Relation}, \emph{ApJ}, {\bf 604}, L89.
\bibitem{9} T.M. Heckman, 1980, \emph{An optical and radio survey of the nuclei of bright galaxies - Activity in normal galactic nuclei}, \emph{A\&A}, {\bf 87}, 152.
\bibitem{10} V.E. Karachentseva, S.N. Mitronova, O.V. Melnyk \& I.D. Karachentsev, 2010, \emph{Catalog of isolated galaxies selected from the 2MASS survey}, \emph{AstBu}, {\bf 65}, 1.
\bibitem{11} S., Kaspi, P.S., Smith, H., Netzer, D., Maoz, B.T., Jannuzi, U., Giveon, 2000, \emph{Reverberation Measurements for 17 Quasars and the Size-Mass-Luminosity Relations in Active Galactic Nuclei}, \emph{ApJ}, {\bf 533}, 631.
\bibitem{12} G. Kauffmann, T.M. Heckman, C. Tremonti, J. Brinchmann, S. Charlot, S.D.M. White, S.E. Ridgway, J. Brinkmann, M. Fukugita, P.B. Hall, Z. Ivezi\'c, G.T. Richards, \& D.P. Schneider, 2003, \emph{The host galaxies of active galactic nuclei}, \emph{MNRAS}, {\bf 346}, 1055.
\bibitem{13} L.J. Kewley, B. Groves, G. Kauffmann \& T. Heckman, 2006, \emph{The host galaxies and classification of active galactic nuclei}, \emph{MNRAS}, {\bf 372}, 961.
\bibitem{14} L.J. Kewley, M.A. Dopita, R.S. Sutherland, C.A. Heisler \& J. Trevena, 2001, \emph{Theoretical Modeling of Starburst Galaxies}, \emph{ApJ}, {\bf 556}, 121.
\bibitem{15} S. Komossa \& D. Xu, 2007, \emph{Narrow-line seyfert 1 galaxies and the M$_{BH}$-$\sigma$ relation}, \emph{ApJ}, {\bf 667}, L33.
\bibitem{16} S. B. Kraemer, I. M. George, D. M. Crenshaw \& J. R. Gabel, 2004, \emph{On the relationship between the optical emission-line and x-ray luminosities in seyfert 1 galaxies}, \emph{ApJ}, {\bf 607}, 794.
\bibitem{17} J.H. Krolik, \emph{in ``Active Galactic Nuclei''}, Princeton University Press, 1999.
\bibitem{18} K. Matsuoka, T. Nagao, A. Marconi, R. Maiolino, \& Y. Tanigushi, 2011, \emph{The mass-metallicity relation of SDSS quasars}, \emph{A\&A}, {\bf 527}, 100.
\bibitem{19} C.J. Miller, R.C. Nichol, P.L. G\'omez, A.M. Hopkins, \& M. Bernardi, 2003, \emph{The Environment of Active Galactic Nuclei in the Sloan Digital Sky Survey}, \emph{ApJ}, {\bf 597}, 142.
\bibitem{20} D.E. Osterbrock, 1989, \emph{Astrophysics of Gaseous Nebulae and Active Galactic Nuclei}, University Science Books.
\bibitem{21} D.E. Osterbrock, 1970, \emph{Abundances of the Elements in Caseous Nebulae}. \emph{QJRAS}, {\bf 11}, 199.
\bibitem{22} K. A. Pounds, C. Done, \& J. P. Osborne, 1995, \emph{RE 1034+39: a high-state Seyfert galaxy?}, \emph{MNRAS}, {\bf 277}, L5.
\bibitem{23} O. Shemmer, H. Netzer, R. Maiolino, E. Oliva, S. Croom, E. Corbett, \& L. di Fabrizzio, 2004, \emph{Near-Infrared Spectroscopy of High-Redshift Active Galactic Nuclei. I. A Metallicity-Accretion Rate Relationship}, \emph{ApJ}, {\bf 614}, 547.
\bibitem{24} T. Storchi Bergmann, 1991, \emph{On the ratio (N II)/H-alpha in the nucleus of Seyfert 2 and LINER galaxies}. \emph{MNRAS}, {\bf 249}, 404.
\bibitem{25} T. Storchi Bergmann, \& M.G. Pastoriza, 1989, \emph{On the metal abundance of low-activity galactic nuclei}. ApJ, 347, 195.
\bibitem{26} J.P. Torres-Papaqui, R. Coziol, H. Andernach, J.M. Islas-Islas, R.A. Ortega-Minakata, D.M. Neri-Larios, \& I. Plauchu-Frayn, 2011, \emph{LLAGNs in dense galactic environments: evidence of dying quasars in massive early type galaxies}, \emph{MNRAS}, submitted.
\bibitem{27} S. Veilleux, \& D.E. Osterbrock, 1987, \emph{Spectral classification of emission-line galaxies}, \emph{ApJS}, {\bf 63}, 295.
\bibitem{28} D.W. Weedman, 1986, \emph{Quasar Astronomy}, Cambridge University Press.
\end{thebibliography}
\end{document}